\documentclass[aps,pra,twocolumn,groupedaddress,showpacs,amsmath,amsfonts,floatfix]{revtex4}
\usepackage{graphicx}
\usepackage{bm}
\usepackage{color}

\begin{document}
\title{Defect State Density and Orbital Localization in a-Si:H/c-Si \\ Heterojunction and the Role of H}

\author{Reza Vatan Meidanshahi}
\email{rvatanme@asu.edu}
\author{Stephen M. Goodnick}
\author{Dragica Vasileska}
\affiliation{School of Electrical, Computer and Energy Engineering, Tempe, AZ, 85281, USA}

\date{\today}

\begin{abstract} 
In this paper, we explore the effect of H and its bonding configurations on the defect state density and orbital localization of hydrogenated amorphous Si (a-Si:H)/crystalline Si (c-Si) heterostructures using density functional theory (DFT) studies of model interfaces between amorphous silicon (a-Si)/a-Si:H and c-Si. To model the atomic configuration of a-Si on c-Si, melting and quenching simulations were performed using classical molecular dynamics (MD). Different hydrogen contents were inserted into the a-Si in different bonding configurations followed by DFT relaxation to create the stable structures of a-Si:H representative of hydrogenated a-Si on crystalline Si surfaces. In contrast to crystalline heterojunctions (where the interface density is a maximum at the interface), we find that, in the most energetically stable configurations of H atoms, the defect state density is relatively low at the interface and maximum at the middle of a-Si layer. Our structural analysis shows that in these configurations, H atoms do not necessarily bond to dangling bonds or to interface atoms. However, they are able to significantly change the atomic structure of the heterostructure and consequently decrease the density of defect states and orbital localization at the a-Si layer and more significantly at the interface of a-Si/c-Si. The general form of the modeled defect state distribution demonstrates the passivating role of a-Si:H on c-Si substrates.
\end{abstract}

\pacs{Valid PACS appear here}

\maketitle
\section{\label{sec:intro}Introduction}

By deposition of a thin layer of a-Si:H on a high quality c-Si substrate, the a-Si:H/c-Si heterojunction solar cells (HIT cells) have achieved the world efficiency record of 26.6\% for n-type Si wafers \cite{yoshikawa2017silicon}. Key to the success of HIT devices is using a a-Si:H layer as a passivating and semiconducting film on top of c-Si \cite{de2012high, yablonovitch1985720, peter2008physics}. The a-Si:H layer reduces the recombination rate of photo-generated carriers by passiviating defect states at the surface of c-Si and controls the photo-generated carriers transport by creating a proper band bending (band offset) at the interface. Engineering these two electronic properties (defect states distribution and band bending) is extremely important to future improve the efficiency of HIT cells. H incorporation in different conditions during plasma enhanced chemical vapor deposition of a-Si:H offer a unique way to engineer the mentioned electronic properties \cite{stuckelberger2014hydrogenated, schulze2011band, brown1997electronic, fantoni2001influence, gall1997spectral, schmidt2007physical, van2012physics}. Due to different fabrication conditions, different H bonding configurations might be generated in a-Si:H layer and at the interface. A better understanding of these H bonding configurations on the electronic structure of a-Si:H/c-Si heterojunctions will aim to improve the device performance. In the current work, we present our first principle study on the electronic properties of a-Si:H/c-Si(001) heterojunction containing different H concentrations with different bonding configurations.

Despite the wide range of literature concerning the simulation of the amorphous Si phase, both pure and hydrogenated, either from first principles and by means of semiempirical approaches, we are not aware of many direct simulations of the a-Si:H/c-Si electronic structure, particularly considering different H concentrations and bonding configurations. In an early work,  Peressi et al. have investigated the effect of defects in the electronic properties of a-Si/c-Si \cite{peressi2001role}. Tosolini et al. \cite{tosolini2004atomic} studied the atomic structure of the interface between a-Si:H and c-Si using combined tight-bonding molecular dynamics simulations and first principle calculations. In another work, Nolan et al. \cite{nolan2012surface} investigated the electronic property changes in a-Si:H and c-Si of a-Si:H/c-Si junction which are perturbed by interface formation on three unreconstructed silicon surfaces, namely (100), (110) and (111). Recently, by combining magnetic resonance measurement and DFT calculations, George et al. \cite{george2013atomicf} studied the atomic structure of interface states of a-Si:H/c-Si heterojunction. In these prior works, in general, the authors mainly focused on the atomic structure and electronic properties of a-Si:H/c-Si heterostructures, while the relation between these properties and the H concentration in different bonding configurations was not fully explored. The main new contribution of the present work compared to previous studies is threefold: (i) a number of different H configurations with differing concentrations were studied, providing useful insight on the fabrication of high quality a-Si:H/c-Si heterojunctions;  (ii) the density of defect states (bandgap states) along the growth direction of the a-Si:H/c-Si heterojunction were calculated, which can be utilized in solar simulation of the electrical behavior of HIT cells, in order to more accurate model their behavior and better engineer of their performance; (iii) orbital localization of the a-Si:H/c-Si heterojunctions were studied, which provides important information that can be used in modeling the recombination rate and effective surface recombination velocity, which in turn strongly affects the efficiency of the HIT solar cells. 

In this paper, we present the results of our first principle calculations performed on a-Si:H/c-Si heterostructure models obtained from Si (100) surfaces. These models were initially generated from partially melting and quenching part of the Si on a Si surface using MD simulation and optimized for different H bonding configurations using DFT methods. The electronic properties of the simulated structures are calculated to elucidate the sensitivity on H concentration and bonding to the defect state density and orbital localization in the a-Si:H/c-Si heterostructure. Our results show that the hydrogen amount and configuration play a key role in the electronic properties.

\section{Method}
\subsection{Technical Details}

Melting and quenching process was simulated on c-Si structures to generate a structural model of a-Si:H/c-Si, as the starting atomic structure for the DFT calculations. We used the LAMMPS molecular dynamic code \cite{plimpton1995fast} to simulate the melting and quenching process, as we have described previously \cite{meidanshahi2019electronic}. Si atom interactions were described by the Tersoff interatomic potential \cite{tersoff1989modeling}, with cut-offs of 2.7   \AA{} (taper) and 3.0   \AA{} (maximum); this interatomic potential has been successfully applied for modeling Si based amorphous structures, as discussed elsewhere \cite{meidanshahi2019electronic, ohira1994molecular}. The full optimization of all the structures obtained from MD were carried out at the DFT-level as implemented in the Quantum Espresso 5.2.1 software package \cite{giannozzi2009quantum}. The Broyden-Fletcher-Goldfarb-Shanno (BFGS) quasi-Newton algorithm, based on the trust radius procedure, was employed as the optimization algorithm for the relaxed structures.

The Perdew-Burke-Ernzerhof (PBE)  \cite{perdew1996generalized} exchange-correlation functional was used in both the ionic relaxation and the electronic structure calculations using periodic boundary conditions. The core and valence electron interactions were described by the Norm-Conserving Pseudopotential function. Unless otherwise stated, an energy cutoff of 12 Ry was employed for the plane-wave basis set and a 4$\times$4$\times$4 k-point mesh is used with the Monkhorst-Pack grid method for the Brillouin-zone sampling in all calculations. A Gaussian smearing and fixed method was applied to determine the band occupations and electronic density of states for odd and even number of H atoms in the supercell, respectively.

\subsection{Generation of a-Si:H/c-Si Structures}

For modeling a-Si:H/c-Si heterostructures, we used periodically repeated supercells containing one interface for each phase and, therefore, two interfaces. We simulated the most common interface, which is that of a-Si:H deposited on a (001) c-Si substrate. To model an a-Si/c-Si interface, a tetragonal 001 supercell containing 256 Si atoms was cleaved from a c-Si structure (diamond lattice) with the experimental lattice constant fixed at 5.43   \AA{} ($a_0$). The lateral dimensions of the supercell were fixed at a=b=2$a_0$ and c=8$a_0$. The longitudinal dimension was chosen according to the experimental density of a-Si:H for given H concentrations \cite{custer1994density, smets2003vacancies}. Once the supercell is determined, the melting and quenching process was performed on the 8 upper layers of the supercell within the constant volume, constant temperature ensemble, while maintaining the c-Si portion in equilibrium at 300 K. In practice, we start with a melting simulation at 3000 K for 10 ps using a time step of 0.1 ps. The structure was rapidly quenched close to room temperature (300 K) using a cooling rate of 6$\times$10\textsuperscript{12} K/s. The cooling rate here is consistent with previously reported cooling rates \cite{ishimaru1997generation, stich1991amorphous, jarolimek2009first}, varying between 10\textsuperscript{11} and 10\textsuperscript{15} K/s. This cooling rate allowed the formation of a disordered a-Si phase interfaced with a crystalline phase. After quenching, the complete structure was annealed for 25 ps at 300 K. Finally, the obtained interface was fully optimized by DFT, as discussed above.

Hydrogenated a-Si structures were obtained by adding H atoms step-by-step to the amorphous constituent of the a-Si/c-Si structure resulting from the MD quenching and DFT optimization above, as described in our previous work \cite{meidanshahi2019electronic}. In each step, one H atom is added to the most stable configuration obtained from the previous step. In order to obtain stable configurations at different H concentrations (i.e. each step), we sample all possible configurations of adding a H atom to each Si atom in the a-Si phase.  The number of Si atoms corresponding to the a-Si phase was 128, and, therefore, for each H addition, we generated 128 different configurations by bonding H to a given Si atom. After optimizing all the configurations using DFT, the configuration with the lowest energy is taken as the most stable one at a given H concentration, and used for further calculations.

\section{Results}
\subsection{Atomic and Electronic Structure}
We first analyzed the atomic structure of the supercells generated by the quenching and DFT optimization discussed in the previous section. The top panel of Fig. 1 (a) shows a ball-and-stick representation of the atomic structure of the a-Si/c-Si supercell.  The supercell contains 256 Si atoms with 128 Si atoms in the a-Si phase. After repeating the 3D supercell, we found ten floating bonds and four dangling bonds (Table I) per supercell using a cutoff Si-Si bond length of 2.58   \AA{}, which is 10\% longer than experimental Si-Si bond length (2.35   \AA{}). Dangling (floating) bonds are missing (extra) chemical bonds of a Si atom from ideally four chemical bonds.

\begin{figure*}[tb]
\includegraphics[width=133mm]{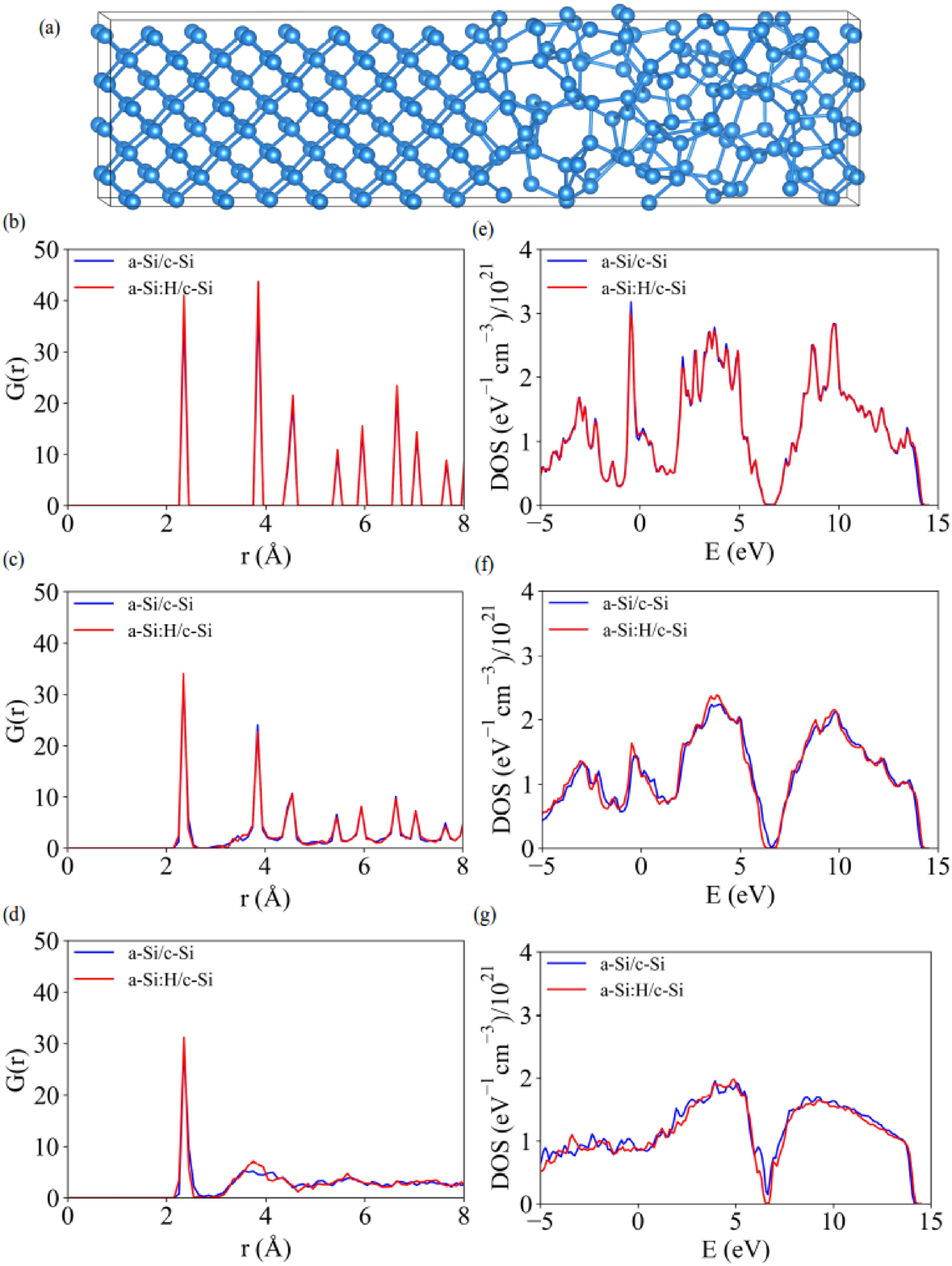}
\caption{ \label{fig1} Top panel: a) the atomic structure of the simulated a-Si/c-Si supercell. Left panel: the layer resolved Si-Si RDF of b) c-Si, c) the interface and d) a-Si. Right panel: the projected Density of States (PDOS) plot of e) c-Si, f) the interface and g) a-Si of a-Si/c-Si in comparison with those of a-Si:H/c-Si (5.88 H at\%).
}\end{figure*}

The structure mostly displays stable 5, 6 or 7 folded rings, and there are no large voids or holes inside. The average Si-Si bond length is 2.37   \AA{} with an rms deviation of 0.04   \AA{} while the average Si-Si-Si bond angle is 107.78$^{\circ}$ with an rms deviation of 12.36$^{\circ}$, which is very close to the upper end of experimental angle distribution of 9-11$^{\circ}$ \cite{pedersen2017optimal}. For reference, the Si-Si bond length is 2.35   \AA{} and Si-Si-Si bond angle is 109.47$^{\circ}$ in the crystalline form of Si. The excess energy of the structure is 0.19 eV/atom which is close to the upper end of experimental excess energy range of 0.07-0.15 eV/atom for a-Si \cite{pedersen2017optimal}. As mentioned earlier, H atoms were gradually added to a-Si/c-Si supercell in different bonding configurations to obtain a-Si:H/c-Si supercells. The average bond length of Si-H bonds in optimized a-Si:H/c-Si structure is 1.5   \AA{}, which is close to the experimental value \cite{lide2004crc}. Table I illustrates the number of dangling bonds (DBs) and floating bonds (FBs) per supercell in the most stable a-Si:H/c-Si heterostructures for different atomic percent of H concentration. As is clearly seen, the number of DBs and FBs per supercell reduces as the H concentration increases. This observation is consistent with the common belief that H incorporation in the growth of a-Si results in passivating DBs and FBs. Interestingly, the number of FBs decreases more rapidly than that of DBs by adding H atoms to the supercells, thus indicating that FBs cause more instability compared to DBs. All DBs and FBs in the a-Si:H/c-Si heterostructure disappear when the concentration of H reaches 5.88\% in the supercell. Therefore, for H bonding configurations study, we only focus on DFT results obtained from the structure with a H concentration of 5.88\%.

In order to check the validity of our atomic structure, we compute the region resolved Si-Si radial distribution function (RDF) of the heterostructure along the growth direction z. Here, a region is defined as a 3D region with the dimensions of the supercell a=b=c=2$a_0$ centered in a given phase (a-Si, c-Si and interface). The RDF gives the probability of finding two atoms in a structure separated by a distance r. Since the RDF is experimentally measured, it provides a strong tool for checking the validity of computationally created structures.

\begin{table}[tb]
\caption{The number of dangling bonds (Num.DB) and floating bonds (Num.FB) in the a-Si:H/c-Si heterostructure per supercell for different atomic percent of H concentration (H at\%).}
\label{comparison_table1}
\begin{tabular}{lc*{8}c}
\hline
\hline
H at\%\;\;\;\; & Num.DB\;\; & Num.FB\\
\hline
\;\;\;0 & 4 & 10  \\
\;0.78 & 4 & 5  \\
\;1.54 & 4 & 2  \\
\;2.29 & 3 & 2  \\
\;3.03 & 2 & 2  \\
\;3.76 & 3 & 0  \\
\;4.47 & 2 & 0  \\
\;5.18 & 0 & 1  \\
\;5.88 & 0 & 0  \\
\hline
\hline
\end{tabular}
\end{table}

The left panel of Fig. 1 displays the computed region resolved Si-Si RDF obtained from the modeled supercells. We expect to observe sharp peaks in the RDF for crystalline materials due to periodic atomic structure, while broadened and coalesced peaks appear in amorphous materials due to the presence of weak short range order and the lack of long range order. Consistent with this argument, we obtained very sharp peaks in the RDF of the crystalline region (Fig. 1b), and broadened peaks for the RDF of the amorphous region (Fig. 1d). Although the amplitude of the RDF peaks does not give quantitative information because of the statistical noise due to the limited size of the supercell, the RDF peak location can provide valid information for comparison purposes. The peak locations in the computed RDFs are consistent with the peak locations in the measured RDFs for both the case of bulk crystalline and amorphous Si \cite{laaziri1999high, bellisent1989structure}.  In addition, the RDF at the interface (Fig. 1c) gives us a visual indication of the transition between the amorphous and crystalline regions. The region resolved RDFs obtained for a-Si:H/c-Si compared to similar ones for a-Si/c-Si show a narrowing of the peaks indicating of increased atomic ordering. This atomic order enhancement is particularly clear in the second RDF peak of a-Si:H in comparison with that of a-Si.

As a result of the atomic structure differences, c-Si and a-Si or a-Si:H have different electronic structure. The c-Si electronic structure contains three main characteristic peaks \cite{chelikowsky1974electronic, ley1972x} in the valence band and a clear energy gap between valence and conduction bands, with no midgap states in the energy gap. In contrast to c-Si, the a-Si or a-Si:H electronic structure contains two main characteristic peaks \cite{singh1981influence} in the valence band and the well known tail and midgap states in the forbidden gap of its crystalline counterpart. These band tail states and midgap states are highly sensitive to the incorporation of hydrogen in the fabrication of a-Si. The aforementioned electronic properties are related to c-Si and a-Si(a-Si:H) in the bulk phase, but here we compared them with the computed projected density of states (PDOS) in the crystalline and amorphous region of the modeled supercell in order to further verify the reliability of the generated structures. The PDOSs were computed in a middle section of the c-Si and a-Si or a-Si:H supercell, at the maximum distance from the interfaces. The right panel of Fig. 1 shows the computed PDOSs. For the c-Si region (Fig. 1e), the PDOS shows the three characteristic peaks of c-Si bulk and a clear separation between the valence and conduction bands. On the other hand, the PDOS of a-Si or a-Si:H (Fig. 1g) shows a deviation from the c-Si PDOS by converting three to two characteristic peaks and by the appearance of valence and conduction band tails and midgap states due to the presence of undercoordinated Si and strained Si-Si bonds \cite{khomyakov2011large}. As seen, the PDOS in the c-Si regions has the typical form of bulk c-Si and the PDOS in the a-Si regions is typical of bulk a-Si. On the other hand, our computed PDOS in the interface region (Fig. 1f) contains a combination of electronic properties of c-Si and a-Si(a-Si:H). While the conduction band of interface PDOS shows three characteristic peaks coming from the c-Si part, the peaks are broadened and collapsed due to the presence of the a-Si part. The bandgap of the interface PDOS contains some midgap states, but not as high as the PDOS of a-Si or a-Si:H.

\subsection{Defect States and Orbital Localization with Different H Concentration}

After confirming the reliability of our structures in the previous section, we performed electronic structure calculations on a-Si:H/c-Si heterostructure on the most stable H bonding configurations, to understand the role of H in the electronic properties. Fig. 2a shows the total electronic density of states (EDOS) of unhydrogenated a-Si/c-Si in comparison with that of a-Si:H/c-Si in its most stable configuration, for a H concentration of 5.88\%. We observe that the effect of hydrogenation is small on the EDOS of the overall valence band and conduction band structure of a-Si. However, looking in more detail at the EDOS close to the valence and conduction band edge, the effect of adding H to the a-Si structure noticeably changes the electronic structure around the band gap.

\begin{figure}[tb]
\includegraphics[width=83mm]{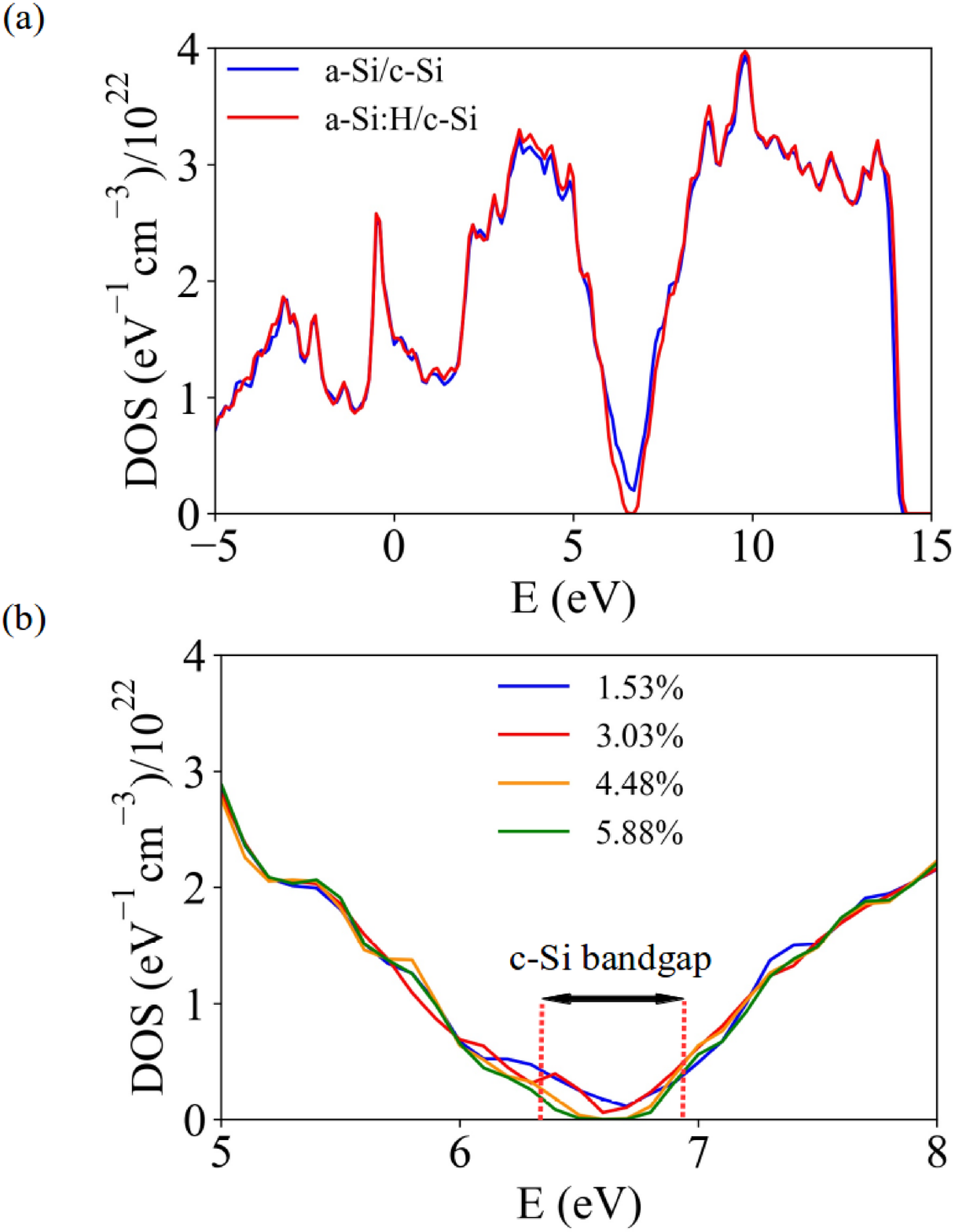}
\caption{ \label{fig2} a) The total EDOS of a-Si/c-Si in comparison to that of a-Si:H/c-Si and b) the EDOS of a-Si:H/c-Si, close to the band gap, for different H concentrations. The numbers assigned to each line correspond to the H atomic percent in the a-Si:H layer.
}\end{figure}
 
Zeroing in on the details near the band edges, Fig. 2b shows the EDOS of an a-Si:H/c-Si structure with different H concentrations for their most stable configurations. It is apparent that, as the H concentration increases, the density of midgap states decreases. Interestingly, the rate of reduction of the midgap states is greater for low H concentrations compared to the rate at higher H concentration. Structural analysis indicates that this decrease in the most stable configuration is not only due to H bonding to Si dangling bonds, but also due to H bonding to Si under high bond length or strain angle. This finding is not unexpected, since Si-Si strained bonds are found to be one of the main sources of midgap and band tail states \cite{meidanshahi2019electronic, khomyakov2011large}, as well as a main source of DBs and FBs due to a high degree of disorder in the a-Si network. Consistent with this argument, we find that in some cases, H atom addition to the Si-Si strained bonds results in non-locally removing DBs and FBs which are considerably further away from the added H location, as we found previously in bulk a-Si \cite{meidanshahi2019electronic}.  Looking in more detail at the EDOS for a-Si:H/c-Si compared to a-Si/c-Si (Fig. 2a), we observe that the reduction in midgap states due to H passivation is associated with an increase in the density of states at the main valence and conduction bands as a consequence of the conservation theorem for the number of eigenstates \cite{khomyakov2011large}. Despite all these changes in the electronic structure of a-Si/c-Si due to H addition, the general form of the electronic structure is preserved for all H concentrations being considered. Neither the exchange-correlation functional or the size of the supercell used here are sufficient to precisely determine the density and position of the midgap and tail states, but our calculations still offer a valid model to predict and analyze the basic changes in the electronic structure due to H addition for potential comparison with experiment. 

As mentioned in the introduction, defect induced states in semiconductors are a critical factor affecting the efficiency of optoelectronic devices such as solar cells. These states play a key role in carrier recombination, carrier trapping, and carrier transport \cite{shi2011roles, nowotny2008titanium, nowotny2008defect, janet2010heterogeneous, hsiao2010electron, li2011applications, duan2015identification, liu2016suppress, muralidharan2015kinetic}. Since the atomic structure changes dramatically going from c-Si to a-Si along the growth direction, it is expected that the density of localized defect states and the orbital localization of the electronic states also change along this direction. Therefore, we also investigate how the density of defect states and the orbital localization changes as the phase gradually changes from c-Si to a-Si, and as H is added to the system.

In several first principle studies, the density of defect states has been calculated by projecting the density of midgap states along simulated supercells \cite{kunstmann2017localized, ngwenya2011defect, santos2014atomistic}. We use the same approach in order to calculate the total density of defect states. The band-gap region is calculated by performing DFT calculations on a c-Si supercell with the same size of the modeled a-Si/c-Si supercell. The band gap region extends from 6.27 eV to 6.90 eV, which is less than experimental value due to the well-known bandgap underestimation problem of conventional DFT. Fig. 3 indicates the projected density of states in this band gap range along the z direction of the a-Si:H/c-Si supercell with different stable concentrations of H.
 
As seen in Fig. 3, for all H concentrations, the density of defect states at the interface is much lower than that at the middle of a-Si:H layer. This observation contradicts the often-invoked picture used in the experimental literature of a-Si/c-Si heterostructures of conventional semiconductor heterostructures where the density of states is highest at the interface due to dissimilar bonding, dangling bonds, etc \cite{meidanshahi2019electronicgap}.  In the case of a-Si:H/c-Si, the transition occurs between an ordered form of Si to a less ordered one, where the maximum disorder is in the bulk of the a-Si where the maximum density of localized bandgap states exists, not at the interface between c-Si and a-Si. As can be seen from Fig. 3, the same observed behavior occurs for all H concentrations; however, the overall density of defect states decreases as more H is inserted to the structure.

\begin{figure}[tb]
\includegraphics[width=83mm]{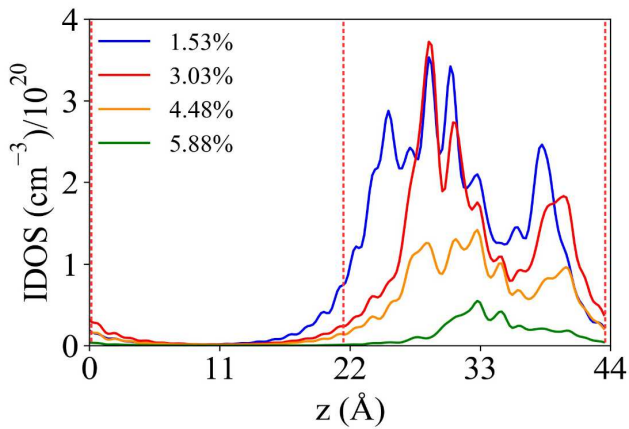}
\caption{ \label{fig3} Integrated density of bandgap states along the z-axis of a-Si:H/c-Si supercell with different H concentrations in the most stable configurations. The vertical red dashed lines show the location of the interfaces in the periodically repeated supercell.  The numbers assigned to each line correspond to the H atomic percent in the a-Si:H layer.
}\end{figure}
 	
Because of the presence of structural defects in the atomic structure of an amorphous material, the mobility of holes and electrons in localized states close to the valence and conduction band edges are much lower in comparison to  nonlocalized extended states that are above/below the mobility edge of the conduction and valence bands, respectively (due to the predominance of hopping conduction \cite{mott2012electronic}). The energy difference between the valence-band mobility edge and the conduction-band mobility edge is the so-called mobility gap. Electronic transport in a-Si and a-Si:H is strongly affected by the carriers residing in these localized states within the mobility gap. Hence, we investigate the orbital localization in a-Si and a-Si:H and the structural defects that are responsible for orbital localization in the mobility gap of a-Si and H addition effects on this localization.

The localization of the Kohn-Sham orbitals was studied in detail for the simulated a-Si:H/c-Si structures using the inverse participation ratio (IPR). The IPR for an eigenstate $\Psi_n$ is given as:

\begin{equation}
IPR_n=\frac{\Sigma_{i=1}^I{a_{ni}^4}}{({\Sigma_{i=1}^I{a_{ni}^2}})^2}
\label{IPR-Eq}
\end{equation}

where $a_{ni}$ is the coefficient of the ith basis set orbital in the nth Kohn-Sham orbital $\Psi_n$ ($\Psi_n=\Sigma_{i=1}^N{a_{ni}\phi_i}$), and N is the total number of basis set orbitals used in the DFT calculations. A higher IPR stands for a higher degree of localization. Fig. 4 shows the calculated IPR of all the Kohn-Sham orbitals obtained via DFT versus their energy for a-Si/c-Si and a-Si:H/c-Si for different H concentrations in their most stable configurations. We observe a relatively sharp transition between the high and low IPR values (localized and extended electronic states) that suggests the presence of mobility edges in both the valence and conduction bands. The general form of the IPR plot is in agreement with previous theoretical and computational studies \cite{legesse2014first, jarolimek2009first}.

\begin{figure}[tb]
\includegraphics[width=83mm]{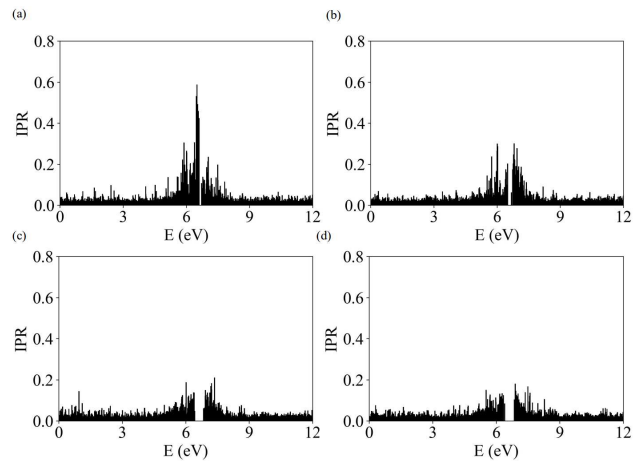}
\caption{ \label{fig4} Orbital localization of a-Si:H/c-Si in different H concentrations: a) 1.56\% b) 3.03\% c) 4.48\% d) 5.88\%.
}\end{figure}
 
From the comparison of the IPR plots shown in Fig. 4, It is obvious that localization of the Kohn-Sham orbitals significantly decreases as hydrogen is added. Since the H atoms add primarily to the strained bond atoms, the decrease in orbital localization is primarily due to removing strained bonds rather than dangling bonds. Localized states strongly influence the effective carrier mobility in terms of scattering centers and hopping conduction \cite{ashcroft1976solid}. Indeed, in the case of a-Si and a-Si:H, it has been experimentally shown that both the electron and hole mobilities are controlled by traps \cite{moore1977electron, marshall1986electron}. Therefore a lower degree of localization is consistent with the observation of higher electron and hole mobility in a-Si:H compared to a-Si experimentally \cite{street1999technology, wagner2008microscopic, hayashi2013role}. We note that H atoms change the orbital localization inside the mobility gap much more than that outside the mobility gap. Therefore, for clearer comparison, we summed all the IPR of the orbitals inside the band gap as an integrated IPR. Fig. 5 shows the integrated IPR of the a-Si:H/c-Si heterostructure as a function of the H concentration.  As seen, the integrated IPR reduces as H is added to the heterstructure, consistent with what is shown in Fig. 4. The rate of reduction is highest at lower H concentrations and then starts to saturate at higher H concentrations. Since the integrated IPR shows the orbital localizations of electronic states in the bandgap, and that localized states in the bandgap are a major source of carrier recombination, this quantity shows that charge recombination rate in a-Si:H/c-Si should decrease with increasing H concentration.

\begin{figure}[tb]
\includegraphics[width=83mm]{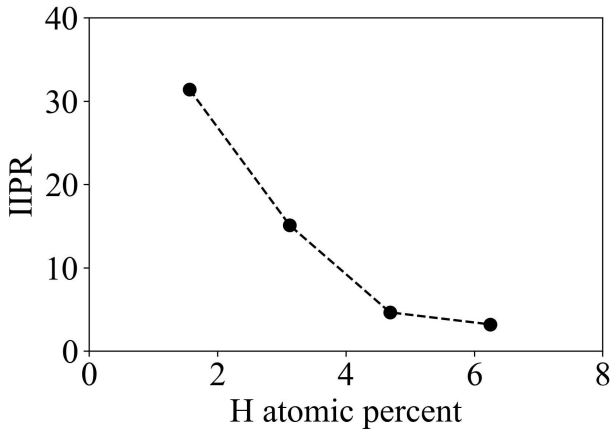}
\caption{ \label{fig5} Integrated IPR (IIPR) of a-Si:H/c-Si for different H concentration.
}\end{figure}

\subsection{Defect States and Orbital Localization with Different H bonding Configurations}

Experimentally, a-Si:H is typically deposited on a c-Si substrate using plasma enhanced chemical vapor deposition (PECVD) under strongly nonequilibrium conditions, where H bond formation is kinetically rather than thermodynamically limited. Hence, different H bonding configurations are expected to occur which are not the energetically lowest state, as considered in the previous section.  In order to investigate how the exact H bonding configuration energetically affects the electronic structure, we examined the electronic structure of the first three most stable configurations of a-Si:H/c-Si heterostructure for a H concentration of 5.88\% (8 H atoms in the amorphous region of the supercell). Fig. 6 shows the EDOS plot close to the band gap of these configurations in comparison with that of unhydrogenated a-Si/c-Si. As shown, no matter in which configuration H atoms are added to the a-Si/c-Si structure, they still reduce the density of midgap states. However, the amount of this reduction depends on the H bonding configurations, and specifically on the energy, with the number of midgap states increasing with increasing excess energy. Structural analysis shows that these two less stable configurations contain DBs and FBs, in contrast to the most stable one, i.e. more structural defects.

\begin{figure}[tb]
\includegraphics[width=83mm]{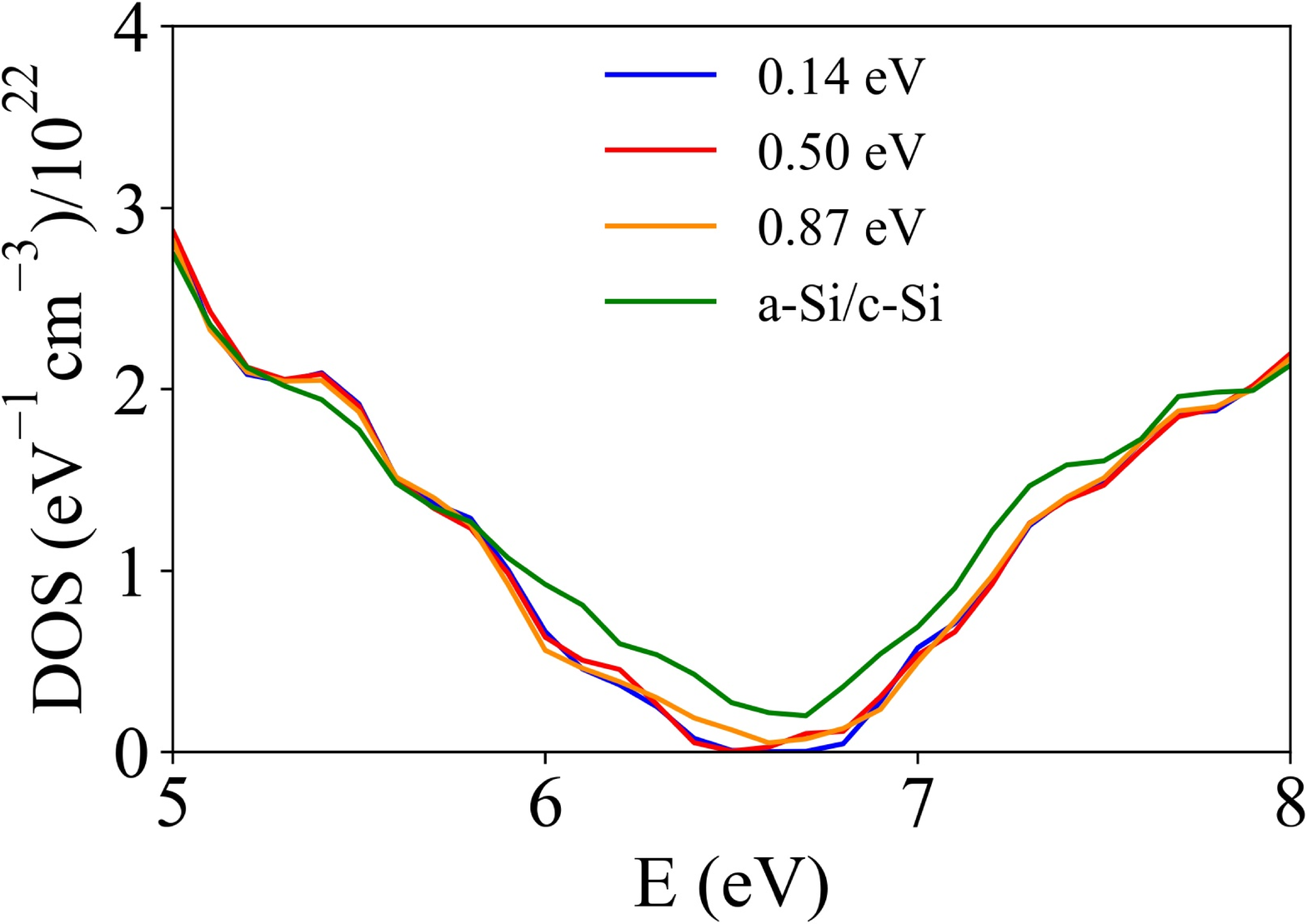}
\caption{ \label{fig6} The density of electronic states as a function of energy of a-Si:H/c-Si structures with 5.88\% H atomic percent in different configurations in comparison with that of a-Si/c-Si. The numbers assigned to each line correspond to the energy of each configuration in eV.
}\end{figure}

Fig. 7 indicates the projected density of states within the c-Si bandgap along the z (growth) direction of the a-Si:H/c-Si for different H bonding configurations in comparison with that of unhydrogenated a-Si/c-Si. Regardless of the H bonding configuration, the density of gap states decreases compared to unhydrogenated case. As expected, the density is zero in the crystalline part of heterojunction and increases monotically until it reaches its maximum inside the a-Si:H layer for all configurations. In the case of a heterojunction between a crystalline and an amorphous form of the same material, the low defect state density at the interface shows that the a-Si:H layer, at least in its most stable configuration, is an effective passivation layer on the top of the c-Si substrate. Thus, there is low defect state density, at the interface that reaches a maximum value in the middle of the a-Si:H layer. Since interface states negatively impact the open circuit voltage in solar cells, the low interface state density helps in explaining the high open circuit voltage in HIT cells. Interestingly, we find that the low density of defect states at the interface is not due to H bonding to the interface atoms. In most of the studied cases, we observe that H atoms preferably bond to atoms in the middle of the a-Si layer. This is due to the higher density of defect sites in the middle of a-Si layer in an a-Si/c-Si heterojunction. It, thus, results in a defect state density reduction not only in the a-Si layer but also at the interface. These results show that H atoms have both local impact in an amorphous network and cause nonlocal changes in the physical and electronic structure. This finding suggests that annealing processes in the presence of H atoms can positively impact the quality of a-Si:H/c-Si heterojunction if H can find its most stable configuration, consistent with recent experimental results \cite{neumuller2018improved}.

\begin{figure}[tb]
\includegraphics[width=83mm]{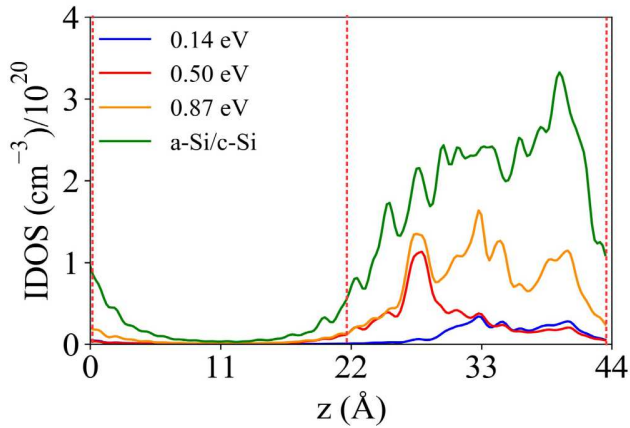}
\caption{ \label{fig7} The local density of defect states for unhydrogenated a-Si/c-Si in comparison with a-Si:H/c-Si with different H bonding configurations of increasing energy. The numbers assigned to each line correspond to the energy of each configuration in eV. The vertical red dashed lines show the location of the interfaces in the periodically repeated supercell.
}\end{figure}

Fig. 8 plots the calculated IPR of all the Kohn-Sham orbitals obtained via DFT versus energy for a-Si:H/c-Si in its first three stable configurations. It is obvious from the figure that regardless of the H bonding configuration, H addition strongly decreases the localization of the electronic states in the a-Si/c-Si heterostructure. Comparing between IPR plots of increasingly energetic configurations of a-Si:H/c-Si heterostructures, the localization of orbitals slowly increases as the energy of the configuration increases. According to the Anderson model, the amount of orbital localization in an amorphous material depends on the amount of disorder in the atomic structure. Therefore, increase in orbital localization can be attributed to the increase in disorder as the energy of the configuration increases. It is worth noting that the distribution of highly localized orbitals also depends on the configuration. In the first stable configuration, most localized states are located in conduction band side, whereas the density of localized states increases in the valence side and midgap region in the second and third configuration. To obtain a more clear vision about orbital localization versus energy of configuration, integrated IPR for different configurations was calculated as described in the previous section. As seen from Fig. 8e, integrated IPR increases as the energy of the configuration increases. 

\begin{figure}[tb]
\includegraphics[width=83mm]{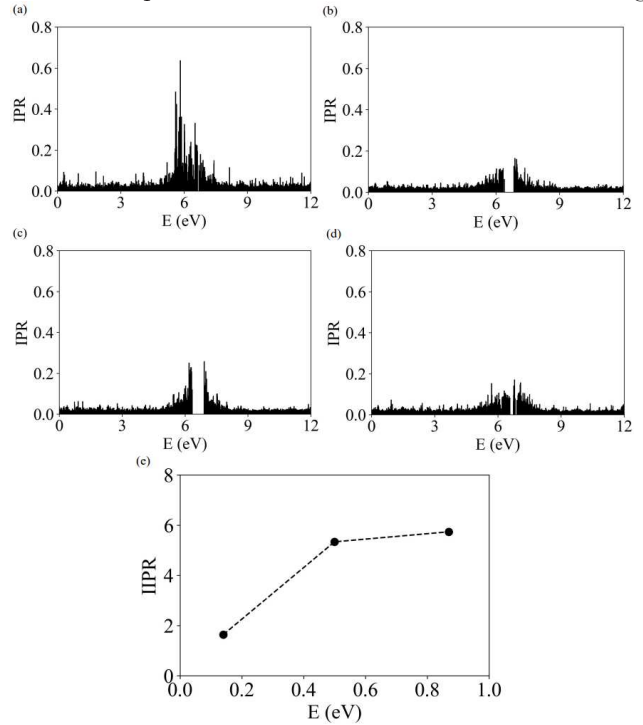}
\caption{ \label{fig8} Orbital localization of a-Si:H/c-Si in comparison to that of a-Si:H/c-Si with H concentrations of 5.88\% for different H bonding configurations as a function of energy: a) a-Si:H/c-Si with no H, b) 0.14 eV c) 0.50 eV, and d) 0.87 eV energy. e) Integrated IPR of a-Si:H/c-Si with 5.88\% H atomic percent for different configurations.
}\end{figure} 

The main result of this section is that when H atoms bind to Si atoms in less energetically stable configurations, they may still reduce the density of bandgap states and orbital localization in the a-Si/c-Si heterostructure.  Moreover, by comparing configurations with different excess energy, we find that the electronic properties are strongly affected by the local H bonding configuration depending on its overall energy.

\section{CONCLUSION}

In conclusion, we employed a combined MD-DFT quenching method to obtain stable configurations of a-Si and a-Si:H. The good agreement between the calculated and experimental RDF and DOS plots, the excess energy and the average bond angle validates the structures used to model the electronic properties of a-Si/c-Si and a-Si:H/c-Si heterostructures. The computed DOS plots obtained from DFT calculations performed on different configurations of a-Si:H/c-Si with different H concentrations indicate that H insertion always decreases the density of states in the forbidden gap, but that the amount of this reduction depends on the H concentration and binding configuration. In contrast to conventional semiconductor heterostructures, the highest density of bandgap defect derived states does not occur at the heterointerface, but rather increases monotonically from the interface to the bulk of the a-Si, due to transition from ordered to disordered material. The interfacial defect density decreases with increasing H concentration which is commensurate with the overall reduction of the defect states throughout the a-Si with H bonding. We found that in the most energetically stable configurations, one has the greatest reduction of midgap states. 

\section{ACKNOWLEDGMENTS}

This material is based upon work primarily supported by the Engineering Research Center Program of the National Science Foundation and the Office of Energy Efficiency and Renewable Energy of the Department of Energy under NSF Cooperative Agreement No.EEC-1041895. Any opinions, findings and conclusions or recommendations expressed in this material are those of the author(s) and do not necessarily reflect those of the National Science Foundation or Department of Energy.

\bibliography{asi_csi}

\begin{thebibliography}{56}
\expandafter\ifx\csname natexlab\endcsname\relax\def\natexlab#1{#1}\fi
\expandafter\ifx\csname bibnamefont\endcsname\relax
  \def\bibnamefont#1{#1}\fi
\expandafter\ifx\csname bibfnamefont\endcsname\relax
  \def\bibfnamefont#1{#1}\fi
\expandafter\ifx\csname citenamefont\endcsname\relax
  \def\citenamefont#1{#1}\fi
\expandafter\ifx\csname url\endcsname\relax
  \def\url#1{\texttt{#1}}\fi
\expandafter\ifx\csname urlprefix\endcsname\relax\def\urlprefix{URL }\fi
\providecommand{\bibinfo}[2]{#2}
\providecommand{\eprint}[2][]{\url{#2}}

\bibitem[{\citenamefont{Yoshikawa et~al.}(2017)\citenamefont{Yoshikawa,
  Kawasaki, Yoshida, Irie, Konishi, Nakano, Uto, Adachi, Kanematsu, Uzu
  et~al.}}]{yoshikawa2017silicon}
\bibinfo{author}{\bibfnamefont{K.}~\bibnamefont{Yoshikawa}},
  \bibinfo{author}{\bibfnamefont{H.}~\bibnamefont{Kawasaki}},
  \bibinfo{author}{\bibfnamefont{W.}~\bibnamefont{Yoshida}},
  \bibinfo{author}{\bibfnamefont{T.}~\bibnamefont{Irie}},
  \bibinfo{author}{\bibfnamefont{K.}~\bibnamefont{Konishi}},
  \bibinfo{author}{\bibfnamefont{K.}~\bibnamefont{Nakano}},
  \bibinfo{author}{\bibfnamefont{T.}~\bibnamefont{Uto}},
  \bibinfo{author}{\bibfnamefont{D.}~\bibnamefont{Adachi}},
  \bibinfo{author}{\bibfnamefont{M.}~\bibnamefont{Kanematsu}},
  \bibinfo{author}{\bibfnamefont{H.}~\bibnamefont{Uzu}}, \bibnamefont{et~al.},
  \bibinfo{journal}{Nature Energy} \textbf{\bibinfo{volume}{2}},
  \bibinfo{pages}{17032} (\bibinfo{year}{2017}).

\bibitem[{\citenamefont{De~Wolf et~al.}(2012)\citenamefont{De~Wolf,
  Descoeudres, Holman, and Ballif}}]{de2012high}
\bibinfo{author}{\bibfnamefont{S.}~\bibnamefont{De~Wolf}},
  \bibinfo{author}{\bibfnamefont{A.}~\bibnamefont{Descoeudres}},
  \bibinfo{author}{\bibfnamefont{Z.~C.} \bibnamefont{Holman}},
  \bibnamefont{and} \bibinfo{author}{\bibfnamefont{C.}~\bibnamefont{Ballif}},
  \bibinfo{journal}{Green} \textbf{\bibinfo{volume}{2}}, \bibinfo{pages}{7}
  (\bibinfo{year}{2012}).

\bibitem[{\citenamefont{Yablonovitch et~al.}(1985)\citenamefont{Yablonovitch,
  Gmitter, Swanson, and Kwark}}]{yablonovitch1985720}
\bibinfo{author}{\bibfnamefont{E.}~\bibnamefont{Yablonovitch}},
  \bibinfo{author}{\bibfnamefont{T.}~\bibnamefont{Gmitter}},
  \bibinfo{author}{\bibfnamefont{R.}~\bibnamefont{Swanson}}, \bibnamefont{and}
  \bibinfo{author}{\bibfnamefont{Y.}~\bibnamefont{Kwark}},
  \bibinfo{journal}{Applied Physics Letters} \textbf{\bibinfo{volume}{47}},
  \bibinfo{pages}{1211} (\bibinfo{year}{1985}).

\bibitem[{\citenamefont{Peter et~al.}(2008)}]{peter2008physics}
\bibinfo{author}{\bibfnamefont{W.}~\bibnamefont{Peter}} \bibnamefont{et~al.},
  \emph{\bibinfo{title}{Physics of Solar Cells: From Principles to New
  Concepts}} (\bibinfo{publisher}{John Wiley \& Sons}, \bibinfo{year}{2008}).

\bibitem[{\citenamefont{St{\"u}ckelberger}(2014)}]{stuckelberger2014hydrogenated}
\bibinfo{author}{\bibfnamefont{M.~E.} \bibnamefont{St{\"u}ckelberger}},
  \bibinfo{journal}{{\'E}cole Polytech F{\'e}d{\'e}rale De Lausanne}
  (\bibinfo{year}{2014}).

\bibitem[{\citenamefont{Schulze et~al.}(2011)\citenamefont{Schulze, Korte,
  Ruske, and Rech}}]{schulze2011band}
\bibinfo{author}{\bibfnamefont{T.}~\bibnamefont{Schulze}},
  \bibinfo{author}{\bibfnamefont{L.}~\bibnamefont{Korte}},
  \bibinfo{author}{\bibfnamefont{F.}~\bibnamefont{Ruske}}, \bibnamefont{and}
  \bibinfo{author}{\bibfnamefont{B.}~\bibnamefont{Rech}},
  \bibinfo{journal}{Physical Review B} \textbf{\bibinfo{volume}{83}},
  \bibinfo{pages}{165314} (\bibinfo{year}{2011}).

\bibitem[{\citenamefont{Brown et~al.}(1997)\citenamefont{Brown, Bittencourt,
  Sebastiani, and Evangelisti}}]{brown1997electronic}
\bibinfo{author}{\bibfnamefont{T.}~\bibnamefont{Brown}},
  \bibinfo{author}{\bibfnamefont{C.}~\bibnamefont{Bittencourt}},
  \bibinfo{author}{\bibfnamefont{M.}~\bibnamefont{Sebastiani}},
  \bibnamefont{and}
  \bibinfo{author}{\bibfnamefont{F.}~\bibnamefont{Evangelisti}},
  \bibinfo{journal}{Physical Review B} \textbf{\bibinfo{volume}{55}},
  \bibinfo{pages}{9904} (\bibinfo{year}{1997}).

\bibitem[{\citenamefont{Fantoni et~al.}(2001)\citenamefont{Fantoni, Vigranenko,
  Fernandes, Schwarz, and Vieira}}]{fantoni2001influence}
\bibinfo{author}{\bibfnamefont{A.}~\bibnamefont{Fantoni}},
  \bibinfo{author}{\bibfnamefont{Y.}~\bibnamefont{Vigranenko}},
  \bibinfo{author}{\bibfnamefont{M.}~\bibnamefont{Fernandes}},
  \bibinfo{author}{\bibfnamefont{R.}~\bibnamefont{Schwarz}}, \bibnamefont{and}
  \bibinfo{author}{\bibfnamefont{M.}~\bibnamefont{Vieira}},
  \bibinfo{journal}{Thin Solid Films} \textbf{\bibinfo{volume}{383}},
  \bibinfo{pages}{314} (\bibinfo{year}{2001}).

\bibitem[{\citenamefont{Gall et~al.}(1997)\citenamefont{Gall, Hirschauer,
  Kolter, and Br{\"a}unig}}]{gall1997spectral}
\bibinfo{author}{\bibfnamefont{S.}~\bibnamefont{Gall}},
  \bibinfo{author}{\bibfnamefont{R.}~\bibnamefont{Hirschauer}},
  \bibinfo{author}{\bibfnamefont{M.}~\bibnamefont{Kolter}}, \bibnamefont{and}
  \bibinfo{author}{\bibfnamefont{D.}~\bibnamefont{Br{\"a}unig}},
  \bibinfo{journal}{Solar energy materials and solar cells}
  \textbf{\bibinfo{volume}{49}}, \bibinfo{pages}{157} (\bibinfo{year}{1997}).

\bibitem[{\citenamefont{Schmidt et~al.}(2007)\citenamefont{Schmidt, Korte,
  Laades, Stangl, Schubert, Angermann, Conrad, and
  Maydell}}]{schmidt2007physical}
\bibinfo{author}{\bibfnamefont{M.}~\bibnamefont{Schmidt}},
  \bibinfo{author}{\bibfnamefont{L.}~\bibnamefont{Korte}},
  \bibinfo{author}{\bibfnamefont{A.}~\bibnamefont{Laades}},
  \bibinfo{author}{\bibfnamefont{R.}~\bibnamefont{Stangl}},
  \bibinfo{author}{\bibfnamefont{C.}~\bibnamefont{Schubert}},
  \bibinfo{author}{\bibfnamefont{H.}~\bibnamefont{Angermann}},
  \bibinfo{author}{\bibfnamefont{E.}~\bibnamefont{Conrad}}, \bibnamefont{and}
  \bibinfo{author}{\bibfnamefont{K.}~\bibnamefont{Maydell}},
  \bibinfo{journal}{Thin Solid Films} \textbf{\bibinfo{volume}{515}},
  \bibinfo{pages}{7475} (\bibinfo{year}{2007}).

\bibitem[{\citenamefont{van Sark et~al.}(2012)\citenamefont{van Sark, Korte,
  and Roca}}]{van2012physics}
\bibinfo{author}{\bibfnamefont{W.}~\bibnamefont{van Sark}},
  \bibinfo{author}{\bibfnamefont{L.}~\bibnamefont{Korte}}, \bibnamefont{and}
  \bibinfo{author}{\bibfnamefont{F.}~\bibnamefont{Roca}},
  \emph{\bibinfo{title}{Physics and technology of amorphous-crystalline
  heterostructure silicon solar cells}} (\bibinfo{publisher}{Springer},
  \bibinfo{year}{2012}).

\bibitem[{\citenamefont{Peressi et~al.}(2001)\citenamefont{Peressi, Colombo,
  and de~Gironcoli}}]{peressi2001role}
\bibinfo{author}{\bibfnamefont{M.}~\bibnamefont{Peressi}},
  \bibinfo{author}{\bibfnamefont{L.}~\bibnamefont{Colombo}}, \bibnamefont{and}
  \bibinfo{author}{\bibfnamefont{S.}~\bibnamefont{de~Gironcoli}},
  \bibinfo{journal}{Physical Review B} \textbf{\bibinfo{volume}{64}},
  \bibinfo{pages}{193303} (\bibinfo{year}{2001}).

\bibitem[{\citenamefont{Tosolini et~al.}(2004)\citenamefont{Tosolini, Colombo,
  and Peressi}}]{tosolini2004atomic}
\bibinfo{author}{\bibfnamefont{M.}~\bibnamefont{Tosolini}},
  \bibinfo{author}{\bibfnamefont{L.}~\bibnamefont{Colombo}}, \bibnamefont{and}
  \bibinfo{author}{\bibfnamefont{M.}~\bibnamefont{Peressi}},
  \bibinfo{journal}{Physical Review B} \textbf{\bibinfo{volume}{69}},
  \bibinfo{pages}{075301} (\bibinfo{year}{2004}).

\bibitem[{\citenamefont{Nolan et~al.}(2012)\citenamefont{Nolan, Legesse, and
  Fagas}}]{nolan2012surface}
\bibinfo{author}{\bibfnamefont{M.}~\bibnamefont{Nolan}},
  \bibinfo{author}{\bibfnamefont{M.}~\bibnamefont{Legesse}}, \bibnamefont{and}
  \bibinfo{author}{\bibfnamefont{G.}~\bibnamefont{Fagas}},
  \bibinfo{journal}{Physical Chemistry Chemical Physics}
  \textbf{\bibinfo{volume}{14}}, \bibinfo{pages}{15173} (\bibinfo{year}{2012}).

\bibitem[{\citenamefont{George et~al.}(2013)\citenamefont{George, Behrends,
  Schnegg, Schulze, Fehr, Korte, Rech, Lips, Rohrm{\"u}ller, Rauls
  et~al.}}]{george2013atomicf}
\bibinfo{author}{\bibfnamefont{B.}~\bibnamefont{George}},
  \bibinfo{author}{\bibfnamefont{J.}~\bibnamefont{Behrends}},
  \bibinfo{author}{\bibfnamefont{A.}~\bibnamefont{Schnegg}},
  \bibinfo{author}{\bibfnamefont{T.}~\bibnamefont{Schulze}},
  \bibinfo{author}{\bibfnamefont{M.}~\bibnamefont{Fehr}},
  \bibinfo{author}{\bibfnamefont{L.}~\bibnamefont{Korte}},
  \bibinfo{author}{\bibfnamefont{B.}~\bibnamefont{Rech}},
  \bibinfo{author}{\bibfnamefont{K.}~\bibnamefont{Lips}},
  \bibinfo{author}{\bibfnamefont{M.}~\bibnamefont{Rohrm{\"u}ller}},
  \bibinfo{author}{\bibfnamefont{E.}~\bibnamefont{Rauls}},
  \bibnamefont{et~al.}, \bibinfo{journal}{Physical review letters}
  \textbf{\bibinfo{volume}{110}}, \bibinfo{pages}{136803}
  (\bibinfo{year}{2013}).

\bibitem[{\citenamefont{Plimpton}(1995)}]{plimpton1995fast}
\bibinfo{author}{\bibfnamefont{S.}~\bibnamefont{Plimpton}},
  \bibinfo{journal}{Journal of computational physics}
  \textbf{\bibinfo{volume}{117}}, \bibinfo{pages}{1} (\bibinfo{year}{1995}).

\bibitem[{\citenamefont{Meidanshahi
  et~al.}(2019{\natexlab{a}})\citenamefont{Meidanshahi, Bowden, and
  Goodnick}}]{meidanshahi2019electronic}
\bibinfo{author}{\bibfnamefont{R.~V.} \bibnamefont{Meidanshahi}},
  \bibinfo{author}{\bibfnamefont{S.}~\bibnamefont{Bowden}}, \bibnamefont{and}
  \bibinfo{author}{\bibfnamefont{S.~M.} \bibnamefont{Goodnick}},
  \bibinfo{journal}{Physical Chemistry Chemical Physics}
  \textbf{\bibinfo{volume}{21}}, \bibinfo{pages}{13248}
  (\bibinfo{year}{2019}{\natexlab{a}}).

\bibitem[{\citenamefont{Tersoff}(1989)}]{tersoff1989modeling}
\bibinfo{author}{\bibfnamefont{J.}~\bibnamefont{Tersoff}},
  \bibinfo{journal}{Physical Review B} \textbf{\bibinfo{volume}{39}},
  \bibinfo{pages}{5566} (\bibinfo{year}{1989}).

\bibitem[{\citenamefont{Ohira et~al.}(1994)\citenamefont{Ohira, Inamuro, and
  Adachi}}]{ohira1994molecular}
\bibinfo{author}{\bibfnamefont{T.}~\bibnamefont{Ohira}},
  \bibinfo{author}{\bibfnamefont{T.}~\bibnamefont{Inamuro}}, \bibnamefont{and}
  \bibinfo{author}{\bibfnamefont{T.}~\bibnamefont{Adachi}},
  \bibinfo{journal}{Solar energy materials and solar cells}
  \textbf{\bibinfo{volume}{34}}, \bibinfo{pages}{565} (\bibinfo{year}{1994}).

\bibitem[{\citenamefont{Giannozzi et~al.}(2009)\citenamefont{Giannozzi, Baroni,
  Bonini, Calandra, Car, Cavazzoni, Ceresoli, Chiarotti, Cococcioni, Dabo
  et~al.}}]{giannozzi2009quantum}
\bibinfo{author}{\bibfnamefont{P.}~\bibnamefont{Giannozzi}},
  \bibinfo{author}{\bibfnamefont{S.}~\bibnamefont{Baroni}},
  \bibinfo{author}{\bibfnamefont{N.}~\bibnamefont{Bonini}},
  \bibinfo{author}{\bibfnamefont{M.}~\bibnamefont{Calandra}},
  \bibinfo{author}{\bibfnamefont{R.}~\bibnamefont{Car}},
  \bibinfo{author}{\bibfnamefont{C.}~\bibnamefont{Cavazzoni}},
  \bibinfo{author}{\bibfnamefont{D.}~\bibnamefont{Ceresoli}},
  \bibinfo{author}{\bibfnamefont{G.~L.} \bibnamefont{Chiarotti}},
  \bibinfo{author}{\bibfnamefont{M.}~\bibnamefont{Cococcioni}},
  \bibinfo{author}{\bibfnamefont{I.}~\bibnamefont{Dabo}}, \bibnamefont{et~al.},
  \bibinfo{journal}{Journal of Physics: Condensed Matter}
  \textbf{\bibinfo{volume}{21}}, \bibinfo{pages}{395502}
  (\bibinfo{year}{2009}).

\bibitem[{\citenamefont{Perdew et~al.}(1996)\citenamefont{Perdew, Burke, and
  Ernzerhof}}]{perdew1996generalized}
\bibinfo{author}{\bibfnamefont{J.~P.} \bibnamefont{Perdew}},
  \bibinfo{author}{\bibfnamefont{K.}~\bibnamefont{Burke}}, \bibnamefont{and}
  \bibinfo{author}{\bibfnamefont{M.}~\bibnamefont{Ernzerhof}},
  \bibinfo{journal}{Physical review letters} \textbf{\bibinfo{volume}{77}},
  \bibinfo{pages}{3865} (\bibinfo{year}{1996}).

\bibitem[{\citenamefont{Custer et~al.}(1994)\citenamefont{Custer, Thompson,
  Jacobson, Poate, Roorda, Sinke, and Spaepen}}]{custer1994density}
\bibinfo{author}{\bibfnamefont{J.}~\bibnamefont{Custer}},
  \bibinfo{author}{\bibfnamefont{M.~O.} \bibnamefont{Thompson}},
  \bibinfo{author}{\bibfnamefont{D.}~\bibnamefont{Jacobson}},
  \bibinfo{author}{\bibfnamefont{J.}~\bibnamefont{Poate}},
  \bibinfo{author}{\bibfnamefont{S.}~\bibnamefont{Roorda}},
  \bibinfo{author}{\bibfnamefont{W.}~\bibnamefont{Sinke}}, \bibnamefont{and}
  \bibinfo{author}{\bibfnamefont{F.}~\bibnamefont{Spaepen}},
  \bibinfo{journal}{Applied physics letters} \textbf{\bibinfo{volume}{64}},
  \bibinfo{pages}{437} (\bibinfo{year}{1994}).

\bibitem[{\citenamefont{Smets et~al.}(2003)\citenamefont{Smets, Kessels, and
  Van~de Sanden}}]{smets2003vacancies}
\bibinfo{author}{\bibfnamefont{A.}~\bibnamefont{Smets}},
  \bibinfo{author}{\bibfnamefont{W.}~\bibnamefont{Kessels}}, \bibnamefont{and}
  \bibinfo{author}{\bibfnamefont{M.}~\bibnamefont{Van~de Sanden}},
  \bibinfo{journal}{Applied physics letters} \textbf{\bibinfo{volume}{82}},
  \bibinfo{pages}{1547} (\bibinfo{year}{2003}).

\bibitem[{\citenamefont{Ishimaru et~al.}(1997)\citenamefont{Ishimaru, Munetoh,
  and Motooka}}]{ishimaru1997generation}
\bibinfo{author}{\bibfnamefont{M.}~\bibnamefont{Ishimaru}},
  \bibinfo{author}{\bibfnamefont{S.}~\bibnamefont{Munetoh}}, \bibnamefont{and}
  \bibinfo{author}{\bibfnamefont{T.}~\bibnamefont{Motooka}},
  \bibinfo{journal}{Physical Review B} \textbf{\bibinfo{volume}{56}},
  \bibinfo{pages}{15133} (\bibinfo{year}{1997}).

\bibitem[{\citenamefont{Stich et~al.}(1991)\citenamefont{Stich, Car, and
  Parrinello}}]{stich1991amorphous}
\bibinfo{author}{\bibfnamefont{I.}~\bibnamefont{Stich}},
  \bibinfo{author}{\bibfnamefont{R.}~\bibnamefont{Car}}, \bibnamefont{and}
  \bibinfo{author}{\bibfnamefont{M.}~\bibnamefont{Parrinello}},
  \bibinfo{journal}{Physical Review B} \textbf{\bibinfo{volume}{44}},
  \bibinfo{pages}{11092} (\bibinfo{year}{1991}).

\bibitem[{\citenamefont{Jarolimek et~al.}(2009)\citenamefont{Jarolimek,
  De~Groot, De~Wijs, and Zeman}}]{jarolimek2009first}
\bibinfo{author}{\bibfnamefont{K.}~\bibnamefont{Jarolimek}},
  \bibinfo{author}{\bibfnamefont{R.}~\bibnamefont{De~Groot}},
  \bibinfo{author}{\bibfnamefont{G.}~\bibnamefont{De~Wijs}}, \bibnamefont{and}
  \bibinfo{author}{\bibfnamefont{M.}~\bibnamefont{Zeman}},
  \bibinfo{journal}{Physical Review B} \textbf{\bibinfo{volume}{79}},
  \bibinfo{pages}{155206} (\bibinfo{year}{2009}).

\bibitem[{\citenamefont{Pedersen et~al.}(2017)\citenamefont{Pedersen,
  Pizzagalli, and J{\'o}nsson}}]{pedersen2017optimal}
\bibinfo{author}{\bibfnamefont{A.}~\bibnamefont{Pedersen}},
  \bibinfo{author}{\bibfnamefont{L.}~\bibnamefont{Pizzagalli}},
  \bibnamefont{and}
  \bibinfo{author}{\bibfnamefont{H.}~\bibnamefont{J{\'o}nsson}},
  \bibinfo{journal}{New Journal of Physics} \textbf{\bibinfo{volume}{19}},
  \bibinfo{pages}{063018} (\bibinfo{year}{2017}).

\bibitem[{\citenamefont{Lide}(2004)}]{lide2004crc}
\bibinfo{author}{\bibfnamefont{D.~R.} \bibnamefont{Lide}},
  \emph{\bibinfo{title}{CRC handbook of chemistry and physics}},
  vol.~\bibinfo{volume}{85} (\bibinfo{publisher}{CRC press},
  \bibinfo{year}{2004}).

\bibitem[{\citenamefont{Laaziri et~al.}(1999)\citenamefont{Laaziri, Kycia,
  Roorda, Chicoine, Robertson, Wang, and Moss}}]{laaziri1999high}
\bibinfo{author}{\bibfnamefont{K.}~\bibnamefont{Laaziri}},
  \bibinfo{author}{\bibfnamefont{S.}~\bibnamefont{Kycia}},
  \bibinfo{author}{\bibfnamefont{S.}~\bibnamefont{Roorda}},
  \bibinfo{author}{\bibfnamefont{M.}~\bibnamefont{Chicoine}},
  \bibinfo{author}{\bibfnamefont{J.}~\bibnamefont{Robertson}},
  \bibinfo{author}{\bibfnamefont{J.}~\bibnamefont{Wang}}, \bibnamefont{and}
  \bibinfo{author}{\bibfnamefont{S.}~\bibnamefont{Moss}},
  \bibinfo{journal}{Physical review letters} \textbf{\bibinfo{volume}{82}},
  \bibinfo{pages}{3460} (\bibinfo{year}{1999}).

\bibitem[{\citenamefont{Bellisent et~al.}(1989)\citenamefont{Bellisent,
  Menelle, Howells, Wright, Brunier, Sinclair, and
  Jansen}}]{bellisent1989structure}
\bibinfo{author}{\bibfnamefont{R.}~\bibnamefont{Bellisent}},
  \bibinfo{author}{\bibfnamefont{A.}~\bibnamefont{Menelle}},
  \bibinfo{author}{\bibfnamefont{W.}~\bibnamefont{Howells}},
  \bibinfo{author}{\bibfnamefont{A.~C.} \bibnamefont{Wright}},
  \bibinfo{author}{\bibfnamefont{T.}~\bibnamefont{Brunier}},
  \bibinfo{author}{\bibfnamefont{R.}~\bibnamefont{Sinclair}}, \bibnamefont{and}
  \bibinfo{author}{\bibfnamefont{F.}~\bibnamefont{Jansen}},
  \bibinfo{journal}{Physica B: Condensed Matter}
  \textbf{\bibinfo{volume}{156}}, \bibinfo{pages}{217} (\bibinfo{year}{1989}).

\bibitem[{\citenamefont{Chelikowsky and
  Cohen}(1974)}]{chelikowsky1974electronic}
\bibinfo{author}{\bibfnamefont{J.~R.} \bibnamefont{Chelikowsky}}
  \bibnamefont{and} \bibinfo{author}{\bibfnamefont{M.~L.} \bibnamefont{Cohen}},
  \bibinfo{journal}{Physical Review B} \textbf{\bibinfo{volume}{10}},
  \bibinfo{pages}{5095} (\bibinfo{year}{1974}).

\bibitem[{\citenamefont{Ley et~al.}(1972)\citenamefont{Ley, Kowalczyk, Pollak,
  and Shirley}}]{ley1972x}
\bibinfo{author}{\bibfnamefont{L.}~\bibnamefont{Ley}},
  \bibinfo{author}{\bibfnamefont{S.}~\bibnamefont{Kowalczyk}},
  \bibinfo{author}{\bibfnamefont{R.}~\bibnamefont{Pollak}}, \bibnamefont{and}
  \bibinfo{author}{\bibfnamefont{D.}~\bibnamefont{Shirley}},
  \bibinfo{journal}{Physical Review Letters} \textbf{\bibinfo{volume}{29}},
  \bibinfo{pages}{1088} (\bibinfo{year}{1972}).

\bibitem[{\citenamefont{Singh}(1981)}]{singh1981influence}
\bibinfo{author}{\bibfnamefont{J.}~\bibnamefont{Singh}},
  \bibinfo{journal}{Physical Review B} \textbf{\bibinfo{volume}{23}},
  \bibinfo{pages}{4156} (\bibinfo{year}{1981}).

\bibitem[{\citenamefont{Khomyakov et~al.}(2011)\citenamefont{Khomyakov,
  Andreoni, Afify, and Curioni}}]{khomyakov2011large}
\bibinfo{author}{\bibfnamefont{P.}~\bibnamefont{Khomyakov}},
  \bibinfo{author}{\bibfnamefont{W.}~\bibnamefont{Andreoni}},
  \bibinfo{author}{\bibfnamefont{N.}~\bibnamefont{Afify}}, \bibnamefont{and}
  \bibinfo{author}{\bibfnamefont{A.}~\bibnamefont{Curioni}},
  \bibinfo{journal}{Physical review letters} \textbf{\bibinfo{volume}{107}},
  \bibinfo{pages}{255502} (\bibinfo{year}{2011}).

\bibitem[{\citenamefont{Shi et~al.}(2011)\citenamefont{Shi, Cui, Liang, Lu,
  Tong, Su, and Liu}}]{shi2011roles}
\bibinfo{author}{\bibfnamefont{J.}~\bibnamefont{Shi}},
  \bibinfo{author}{\bibfnamefont{H.}~\bibnamefont{Cui}},
  \bibinfo{author}{\bibfnamefont{Z.}~\bibnamefont{Liang}},
  \bibinfo{author}{\bibfnamefont{X.}~\bibnamefont{Lu}},
  \bibinfo{author}{\bibfnamefont{Y.}~\bibnamefont{Tong}},
  \bibinfo{author}{\bibfnamefont{C.}~\bibnamefont{Su}}, \bibnamefont{and}
  \bibinfo{author}{\bibfnamefont{H.}~\bibnamefont{Liu}},
  \bibinfo{journal}{Energy \& Environmental Science}
  \textbf{\bibinfo{volume}{4}}, \bibinfo{pages}{466} (\bibinfo{year}{2011}).

\bibitem[{\citenamefont{Nowotny}(2008)}]{nowotny2008titanium}
\bibinfo{author}{\bibfnamefont{J.}~\bibnamefont{Nowotny}},
  \bibinfo{journal}{Energy \& Environmental Science}
  \textbf{\bibinfo{volume}{1}}, \bibinfo{pages}{565} (\bibinfo{year}{2008}).

\bibitem[{\citenamefont{Nowotny et~al.}(2008)\citenamefont{Nowotny, Sheppard,
  Bak, and Nowotny}}]{nowotny2008defect}
\bibinfo{author}{\bibfnamefont{M.~K.} \bibnamefont{Nowotny}},
  \bibinfo{author}{\bibfnamefont{L.~R.} \bibnamefont{Sheppard}},
  \bibinfo{author}{\bibfnamefont{T.}~\bibnamefont{Bak}}, \bibnamefont{and}
  \bibinfo{author}{\bibfnamefont{J.}~\bibnamefont{Nowotny}},
  \bibinfo{journal}{The Journal of Physical Chemistry C}
  \textbf{\bibinfo{volume}{112}}, \bibinfo{pages}{5275} (\bibinfo{year}{2008}).

\bibitem[{\citenamefont{Janet et~al.}(2010)\citenamefont{Janet, Navaladian,
  Viswanathan, Varadarajan, and Viswanath}}]{janet2010heterogeneous}
\bibinfo{author}{\bibfnamefont{C.}~\bibnamefont{Janet}},
  \bibinfo{author}{\bibfnamefont{S.}~\bibnamefont{Navaladian}},
  \bibinfo{author}{\bibfnamefont{B.}~\bibnamefont{Viswanathan}},
  \bibinfo{author}{\bibfnamefont{T.}~\bibnamefont{Varadarajan}},
  \bibnamefont{and}
  \bibinfo{author}{\bibfnamefont{R.}~\bibnamefont{Viswanath}},
  \bibinfo{journal}{The Journal of Physical Chemistry C}
  \textbf{\bibinfo{volume}{114}}, \bibinfo{pages}{2622} (\bibinfo{year}{2010}).

\bibitem[{\citenamefont{Hsiao et~al.}(2010)\citenamefont{Hsiao, Tung, and
  Teng}}]{hsiao2010electron}
\bibinfo{author}{\bibfnamefont{P.-T.} \bibnamefont{Hsiao}},
  \bibinfo{author}{\bibfnamefont{Y.-L.} \bibnamefont{Tung}}, \bibnamefont{and}
  \bibinfo{author}{\bibfnamefont{H.}~\bibnamefont{Teng}}, \bibinfo{journal}{The
  Journal of Physical Chemistry C} \textbf{\bibinfo{volume}{114}},
  \bibinfo{pages}{6762} (\bibinfo{year}{2010}).

\bibitem[{\citenamefont{Li et~al.}(2011)\citenamefont{Li, Crandall, Repins,
  Nardes, and Levi}}]{li2011applications}
\bibinfo{author}{\bibfnamefont{J.~V.} \bibnamefont{Li}},
  \bibinfo{author}{\bibfnamefont{R.~S.} \bibnamefont{Crandall}},
  \bibinfo{author}{\bibfnamefont{I.~L.} \bibnamefont{Repins}},
  \bibinfo{author}{\bibfnamefont{A.~M.} \bibnamefont{Nardes}},
  \bibnamefont{and} \bibinfo{author}{\bibfnamefont{D.~H.} \bibnamefont{Levi}},
  in \emph{\bibinfo{booktitle}{Photovoltaic Specialists Conference (PVSC), 2011
  37th IEEE}} (\bibinfo{organization}{IEEE}, \bibinfo{year}{2011}), pp.
  \bibinfo{pages}{000075--000078}.

\bibitem[{\citenamefont{Duan et~al.}(2015)\citenamefont{Duan, Zhou, Chen, Sun,
  Luo, Song, Bob, and Yang}}]{duan2015identification}
\bibinfo{author}{\bibfnamefont{H.-S.} \bibnamefont{Duan}},
  \bibinfo{author}{\bibfnamefont{H.}~\bibnamefont{Zhou}},
  \bibinfo{author}{\bibfnamefont{Q.}~\bibnamefont{Chen}},
  \bibinfo{author}{\bibfnamefont{P.}~\bibnamefont{Sun}},
  \bibinfo{author}{\bibfnamefont{S.}~\bibnamefont{Luo}},
  \bibinfo{author}{\bibfnamefont{T.-B.} \bibnamefont{Song}},
  \bibinfo{author}{\bibfnamefont{B.}~\bibnamefont{Bob}}, \bibnamefont{and}
  \bibinfo{author}{\bibfnamefont{Y.}~\bibnamefont{Yang}},
  \bibinfo{journal}{Physical Chemistry Chemical Physics}
  \textbf{\bibinfo{volume}{17}}, \bibinfo{pages}{112} (\bibinfo{year}{2015}).

\bibitem[{\citenamefont{Liu et~al.}(2016)\citenamefont{Liu, Stradins, Deng,
  Luo, and Wei}}]{liu2016suppress}
\bibinfo{author}{\bibfnamefont{Y.}~\bibnamefont{Liu}},
  \bibinfo{author}{\bibfnamefont{P.}~\bibnamefont{Stradins}},
  \bibinfo{author}{\bibfnamefont{H.}~\bibnamefont{Deng}},
  \bibinfo{author}{\bibfnamefont{J.}~\bibnamefont{Luo}}, \bibnamefont{and}
  \bibinfo{author}{\bibfnamefont{S.-H.} \bibnamefont{Wei}},
  \bibinfo{journal}{Applied Physics Letters} \textbf{\bibinfo{volume}{108}},
  \bibinfo{pages}{022101} (\bibinfo{year}{2016}).

\bibitem[{\citenamefont{Muralidharan et~al.}(2015)\citenamefont{Muralidharan,
  Vasileska, Goodnick, and Bowden}}]{muralidharan2015kinetic}
\bibinfo{author}{\bibfnamefont{P.}~\bibnamefont{Muralidharan}},
  \bibinfo{author}{\bibfnamefont{D.}~\bibnamefont{Vasileska}},
  \bibinfo{author}{\bibfnamefont{S.~M.} \bibnamefont{Goodnick}},
  \bibnamefont{and} \bibinfo{author}{\bibfnamefont{S.}~\bibnamefont{Bowden}},
  in \emph{\bibinfo{booktitle}{Photovoltaic Specialist Conference (PVSC), 2015
  IEEE 42nd}} (\bibinfo{organization}{IEEE}, \bibinfo{year}{2015}), pp.
  \bibinfo{pages}{1--4}.

\bibitem[{\citenamefont{Kunstmann et~al.}(2017)\citenamefont{Kunstmann,
  Wendumu, and Seifert}}]{kunstmann2017localized}
\bibinfo{author}{\bibfnamefont{J.}~\bibnamefont{Kunstmann}},
  \bibinfo{author}{\bibfnamefont{T.~B.} \bibnamefont{Wendumu}},
  \bibnamefont{and} \bibinfo{author}{\bibfnamefont{G.}~\bibnamefont{Seifert}},
  \bibinfo{journal}{physica status solidi (b)} \textbf{\bibinfo{volume}{254}}
  (\bibinfo{year}{2017}).

\bibitem[{\citenamefont{Ngwenya et~al.}(2011)\citenamefont{Ngwenya, Ukpong, and
  Chetty}}]{ngwenya2011defect}
\bibinfo{author}{\bibfnamefont{T.~B.} \bibnamefont{Ngwenya}},
  \bibinfo{author}{\bibfnamefont{A.}~\bibnamefont{Ukpong}}, \bibnamefont{and}
  \bibinfo{author}{\bibfnamefont{N.}~\bibnamefont{Chetty}},
  \bibinfo{journal}{Physical Review B} \textbf{\bibinfo{volume}{84}},
  \bibinfo{pages}{245425} (\bibinfo{year}{2011}).

\bibitem[{\citenamefont{Santos et~al.}(2014)\citenamefont{Santos, Cazzaniga,
  Onida, and Colombo}}]{santos2014atomistic}
\bibinfo{author}{\bibfnamefont{I.}~\bibnamefont{Santos}},
  \bibinfo{author}{\bibfnamefont{M.}~\bibnamefont{Cazzaniga}},
  \bibinfo{author}{\bibfnamefont{G.}~\bibnamefont{Onida}}, \bibnamefont{and}
  \bibinfo{author}{\bibfnamefont{L.}~\bibnamefont{Colombo}},
  \bibinfo{journal}{Journal of Physics: Condensed Matter}
  \textbf{\bibinfo{volume}{26}}, \bibinfo{pages}{095001}
  (\bibinfo{year}{2014}).

\bibitem[{\citenamefont{Meidanshahi
  et~al.}(2019{\natexlab{b}})\citenamefont{Meidanshahi, Zhang, Zou, Honsberg,
  and Goodnick}}]{meidanshahi2019electronicgap}
\bibinfo{author}{\bibfnamefont{R.~V.} \bibnamefont{Meidanshahi}},
  \bibinfo{author}{\bibfnamefont{C.}~\bibnamefont{Zhang}},
  \bibinfo{author}{\bibfnamefont{Y.}~\bibnamefont{Zou}},
  \bibinfo{author}{\bibfnamefont{C.}~\bibnamefont{Honsberg}}, \bibnamefont{and}
  \bibinfo{author}{\bibfnamefont{S.~M.} \bibnamefont{Goodnick}},
  \bibinfo{journal}{Progress in Photovoltaics: Research and Applications}
  \textbf{\bibinfo{volume}{27}}, \bibinfo{pages}{724}
  (\bibinfo{year}{2019}{\natexlab{b}}).

\bibitem[{\citenamefont{Mott and Davis}(2012)}]{mott2012electronic}
\bibinfo{author}{\bibfnamefont{N.~F.} \bibnamefont{Mott}} \bibnamefont{and}
  \bibinfo{author}{\bibfnamefont{E.~A.} \bibnamefont{Davis}},
  \emph{\bibinfo{title}{Electronic processes in non-crystalline materials}}
  (\bibinfo{publisher}{Oxford university press}, \bibinfo{year}{2012}).

\bibitem[{\citenamefont{Legesse et~al.}(2014)\citenamefont{Legesse, Nolan, and
  Fagas}}]{legesse2014first}
\bibinfo{author}{\bibfnamefont{M.}~\bibnamefont{Legesse}},
  \bibinfo{author}{\bibfnamefont{M.}~\bibnamefont{Nolan}}, \bibnamefont{and}
  \bibinfo{author}{\bibfnamefont{G.}~\bibnamefont{Fagas}},
  \bibinfo{journal}{Journal of Applied Physics} \textbf{\bibinfo{volume}{115}},
  \bibinfo{pages}{203711} (\bibinfo{year}{2014}).

\bibitem[{\citenamefont{Ashcroft et~al.}(1976)\citenamefont{Ashcroft, Mermin
  et~al.}}]{ashcroft1976solid}
\bibinfo{author}{\bibfnamefont{N.~W.} \bibnamefont{Ashcroft}},
  \bibinfo{author}{\bibfnamefont{N.~D.} \bibnamefont{Mermin}},
  \bibnamefont{et~al.}, \emph{\bibinfo{title}{Solid state physics [by] neil w.
  ashcroft [and] n. david mermin.}} (\bibinfo{year}{1976}).

\bibitem[{\citenamefont{Moore}(1977)}]{moore1977electron}
\bibinfo{author}{\bibfnamefont{A.}~\bibnamefont{Moore}},
  \bibinfo{journal}{Applied Physics Letters} \textbf{\bibinfo{volume}{31}},
  \bibinfo{pages}{762} (\bibinfo{year}{1977}).

\bibitem[{\citenamefont{Marshall et~al.}(1986)\citenamefont{Marshall, Street,
  and Thompson}}]{marshall1986electron}
\bibinfo{author}{\bibfnamefont{J.}~\bibnamefont{Marshall}},
  \bibinfo{author}{\bibfnamefont{R.}~\bibnamefont{Street}}, \bibnamefont{and}
  \bibinfo{author}{\bibfnamefont{M.}~\bibnamefont{Thompson}},
  \bibinfo{journal}{Philosophical magazine B} \textbf{\bibinfo{volume}{54}},
  \bibinfo{pages}{51} (\bibinfo{year}{1986}).

\bibitem[{\citenamefont{Street}(1999)}]{street1999technology}
\bibinfo{author}{\bibfnamefont{R.}~\bibnamefont{Street}},
  \emph{\bibinfo{title}{Technology and applications of amorphous silicon}},
  vol.~\bibinfo{volume}{37} (\bibinfo{publisher}{Springer Science \& Business
  Media}, \bibinfo{year}{1999}).

\bibitem[{\citenamefont{Wagner and Grossman}(2008)}]{wagner2008microscopic}
\bibinfo{author}{\bibfnamefont{L.~K.} \bibnamefont{Wagner}} \bibnamefont{and}
  \bibinfo{author}{\bibfnamefont{J.~C.} \bibnamefont{Grossman}},
  \bibinfo{journal}{Physical review letters} \textbf{\bibinfo{volume}{101}},
  \bibinfo{pages}{265501} (\bibinfo{year}{2008}).

\bibitem[{\citenamefont{Hayashi et~al.}(2013)\citenamefont{Hayashi, Li, Ogura,
  and Ohshita}}]{hayashi2013role}
\bibinfo{author}{\bibfnamefont{Y.}~\bibnamefont{Hayashi}},
  \bibinfo{author}{\bibfnamefont{D.}~\bibnamefont{Li}},
  \bibinfo{author}{\bibfnamefont{A.}~\bibnamefont{Ogura}}, \bibnamefont{and}
  \bibinfo{author}{\bibfnamefont{Y.}~\bibnamefont{Ohshita}},
  \bibinfo{journal}{IEEE Journal of Photovoltaics}
  \textbf{\bibinfo{volume}{3}}, \bibinfo{pages}{1149} (\bibinfo{year}{2013}).

\bibitem[{\citenamefont{Neum{\"u}ller et~al.}(2018)\citenamefont{Neum{\"u}ller,
  Sergeev, Heise, Bereznev, Volobujeva, Salas, Vehse, and
  Agert}}]{neumuller2018improved}
\bibinfo{author}{\bibfnamefont{A.}~\bibnamefont{Neum{\"u}ller}},
  \bibinfo{author}{\bibfnamefont{O.}~\bibnamefont{Sergeev}},
  \bibinfo{author}{\bibfnamefont{S.~J.} \bibnamefont{Heise}},
  \bibinfo{author}{\bibfnamefont{S.}~\bibnamefont{Bereznev}},
  \bibinfo{author}{\bibfnamefont{O.}~\bibnamefont{Volobujeva}},
  \bibinfo{author}{\bibfnamefont{J.~F.~L.} \bibnamefont{Salas}},
  \bibinfo{author}{\bibfnamefont{M.}~\bibnamefont{Vehse}}, \bibnamefont{and}
  \bibinfo{author}{\bibfnamefont{C.}~\bibnamefont{Agert}},
  \bibinfo{journal}{Nano Energy} \textbf{\bibinfo{volume}{43}},
  \bibinfo{pages}{228} (\bibinfo{year}{2018}).

\end{thebibliography}

\end{document}